 \documentclass[twocolumn]{aastex63}

\usepackage{natbib}
\usepackage[style=american]{csquotes}
\newcommand{\tess}{\emph{TESS}}

\usepackage{scrextend}
\let\orgautoref\autoref
\renewcommand{\autoref}
        {\def\equationautorefname{Eq.}%
         \def\figureautorefname{Fig.}%
         \def\sectionautorefname{Sect.}%
         \def\subsectionautorefname{Sect.}%
         \def\subsubsectionautorefname{Sect.}%
         \orgautoref}

\usepackage{changepage} % for \begin{adjustwidth}{-Xcm}{} \end{adjustwidth}
\usepackage{rotating}

\revised{February 13, 2023}
\accepted{February 20, 2023}
\published{March 29, 2023}
\usepackage[bulgarian,english]{babel}
\usepackage[T2A]{fontenc}

\shorttitle{A 2:1 Mean-Motion Resonance Super-Jovian pair}
\shortauthors{Bozhilov et al.}

\begin{document}

\title{{A 2:1 Mean-Motion Resonance Super-Jovian pair revealed by TESS, FEROS, and HARPS\footnote{Based on observations collected at the European Organization for Astronomical Research in the Southern Hemisphere under ESO programmes 105.20GX.001, 108.22A8.001, 110.23YQ.001, and MPG programmes 0106.A-9014, 0107.A-9003, 0108.A-9003, 0109.A-9003, 0110.A-9011.}}}

\correspondingauthor{Vladimir Bohzilov}
\email{vbozhilov@phys.uni-sofia.bg}

\author[0000-0002-3117-7197] {Vladimir Bozhilov}
\affiliation{Department
 of Astronomy, Faculty of Physics, Sofia University ``St Kliment Ohridski'', 5 James Bourchier Blvd, BG-1164 Sofia, Bulgaria}

 \author[0000-0003-1507-7230]{Desislava Antonova}
\affiliation{Department
 of Astronomy, Faculty of Physics, Sofia University ``St Kliment Ohridski'', 5 James Bourchier Blvd, BG-1164 Sofia, Bulgaria}

 \author[0000-0002-5945-7975]{Melissa J.\ Hobson}
\affiliation{Max-Planck-Institut für Astronomie,
              Königstuhl 17,
              69117 Heidelberg, Germany}
\affil{Millennium Institute for Astrophysics, Chile}
 
\author[0000-0002-9158-7315]{Rafael Brahm}
\affil{Facultad de Ingeniera y Ciencias, Universidad Adolfo Ib\'{a}\~{n}ez, Av. Diagonal las Torres 2640, Pe\~{n}alol\'{e}n, Santiago, Chile}
\affil{Millennium Institute for Astrophysics, Chile}
\affil{Data Observatory Foundation, Chile}
 
\author[0000-0002-5389-3944]{Andr\'es Jord\'an}
\affil{Facultad de Ingeniera y Ciencias, Universidad Adolfo Ib\'{a}\~{n}ez, Av. Diagonal las Torres 2640, Pe\~{n}alol\'{e}n, Santiago, Chile}
\affil{Millennium Institute for Astrophysics, Chile}
\affil{Data Observatory Foundation, Chile}
 
\author[0000-0002-1493-300X]{Thomas Henning}
\affiliation{Max-Planck-Institut für Astronomie,
              Königstuhl 17,
              69117 Heidelberg, Germany}

\author[0000-0003-3130-2768]{Jan Eberhardt}
\affiliation{Max-Planck-Institut für Astronomie,
              Königstuhl 17,
              69117 Heidelberg, Germany}

\author[0000-0003-3047-6272]{Felipe I.\ Rojas}
\affil{Instituto de Astrof\'isica, Facultad de F\'isica, Pontificia Universidad Cat\'olica de Chile, Chile}
\affil{Millennium Institute for Astrophysics, Chile}

\author[0000-0002-7094-7908]{Konstantin Batygin}
\affil{Division of Geological and Planetary Sciences, California Institute of Technology, Pasadena, CA 91125, USA}

\author[0000-0003-3047-6272]{Pascal Torres-Miranda}
\affil{Instituto de Astrof\'isica, Facultad de F\'isica, Pontificia Universidad Cat\'olica de Chile, Chile}
\affil{Millennium Institute for Astrophysics, Chile}

\author[0000-0002-3481-9052]{Keivan G.\ Stassun}
\affiliation{Department of Physics and Astronomy, Vanderbilt University, Nashville, TN 37235, USA}

\author[0000-0003-3130-2282]{Sarah C. Millholland}
\affiliation{MIT Kavli Institute for Astrophysics and Space Research, Massachusetts Institute of Technology, Cambridge, MA 02139, USA}

\author[0000-0001-6277-9644]{Denitza Stoeva}
\affiliation{Department
 of Astronomy, Faculty of Physics, Sofia University ``St Kliment Ohridski'', 5 James Bourchier Blvd, BG-1164 Sofia, Bulgaria}

\author[0000-0002-5702-5095]{Milen Minev}
\affiliation{Department
 of Astronomy, Faculty of Physics, Sofia University ``St Kliment Ohridski'', 5 James Bourchier Blvd, BG-1164 Sofia, Bulgaria}
 \affiliation{Institute of Astronomy and National Astronomical Observatory, Bulgarian Academy of Sciences, 72 Tsarigradsko shosse Blvd., 1784 Sofia, Bulgaria}
 
\author[0000-0001-9513-1449]{Nestor Espinoza}
\affil{Space Telescope Science Institute, 3700 San Martin Drive, Baltimore, MD 21218, USA}
 
%\author{TESS architects, TESS contributors}

% TESS architects
\author[0000-0003-2058-6662]{George R. Ricker} % grr@space.mit.edu
\affiliation{Department of Physics and Kavli Institute for Astrophysics and Space Research, Massachusetts Institute of Technology, Cambridge, MA 02139, USA}

\author[0000-0001-9911-7388]{David W. Latham}% dlatham@cfa.harvard.edu
\affiliation{Center for Astrophysics \textbar \ Harvard \& Smithsonian, 60 Garden St, Cambridge, MA 02138, USA}

\author[0000-0003-2313-467X]{Diana Dragomir} 
\affiliation{Department of Physics and Astronomy, University of New Mexico, Albuquerque, NM, USA}

\author[0000-0001-9269-8060]{Michelle Kunimoto} 
\affiliation{Department of Physics and Kavli Institute for Astrophysics and Space Research, Massachusetts Institute of Technology, Cambridge, MA 02139, USA}

\author[0000-0002-4715-9460]{Jon M. Jenkins} % jon.jenkins@nasa.gov
\affiliation{NASA Ames Research Center, Moffett Field, CA 94035, USA}

  \author[0000-0002-8219-9505]{Eric B. Ting} 
\affiliation{NASA Ames Research Center, Moffett Field, CA 94035, USA}
 
\author[0000-0002-6892-6948]{Sara Seager} % seager@mit.edu
\affiliation{Department of Physics and Kavli Institute for Astrophysics
and Space Research, Massachusetts Institute of Technology, Cambridge, MA
02139, USA}
\affiliation{Department of Earth, Atmospheric and Planetary Sciences,
Massachusetts Institute of Technology, Cambridge, MA 02139, USA}
\affiliation{Department of Aeronautics and Astronautics, MIT, 77
Massachusetts Avenue, Cambridge, MA 02139, USA}

\author[0000-0002-4265-047X]{Joshua N. Winn} %jnwinn@princeton.edu,
\affiliation{Department of Astrophysical Sciences, Princeton University,
NJ 08544, USA}

\author[0000-0002-4625-8264]{Jesus Noel Villasenor},
\affiliation{Department of Physics and Kavli Institute for Astrophysics
and Space Research, Massachusetts Institute of Technology, Cambridge, MA
02139, USA}

\author[0000-0002-0514-5538]{Luke G. Bouma}
\affiliation{Cahill Center for Astrophysics, California Institute of Technology, Pasadena, CA 91125, USA}

\author[0000-0002-3284-4713]{Jennifer Medina} 
\affiliation{Space Telescope Science Institute, 3700 San Martin Drive, Baltimore, MD 21218, USA}
 
\author[0000-0002-0236-775X]{Trifon Trifonov}
\affiliation{Department
 of Astronomy, Faculty of Physics, Sofia University ``St Kliment Ohridski'', 5 James Bourchier Blvd, BG-1164 Sofia, Bulgaria}
\affiliation{Max-Planck-Institut für Astronomie,
              Königstuhl 17,
              69117 Heidelberg, Germany}

\begin{abstract}
We report the discovery of a super-Jovian 2:1 mean-motion resonance (MMR) pair around the G-type star TIC\,279401253, whose dynamical architecture is a prospective benchmark for planet formation and orbital evolution analysis.
The system was discovered thanks to a single-transit event recorded by the {\em Transiting Exoplanet Survey Satellite} ($\tess$) mission, which pointed to a Jupiter-sized companion with poorly constrained orbital parameters. 
 We began ground-based precise radial velocity (RV) monitoring with HARPS and FEROS within the {\em Warm gIaNts with tEss} (WINE) survey to constrain the transiting body's period, mass, and eccentricity. The RV measurements revealed not one but two massive planets with periods of 76.80$_{-0.06}^{+0.06}$ and 155.3$_{-0.7}^{+0.7}$ days, respectively.
 A combined analysis of transit and RV data yields an inner transiting planet with a mass of 6.14$_{-0.42}^{+0.39}$\,M$_{\rm Jup}$ and a radius of 1.00$_{-0.04}^{+0.04}$\,R$_{\rm Jup}$, and an outer planet with a minimum mass of 8.02$_{-0.18}^{+0.18}$\,M$_{\rm Jup}$, indicating a massive giant pair. A detailed dynamical analysis of the system reveals that the planets are locked in a strong first-order, eccentricity-type 2:1 MMR, which makes TIC\,279401253
 one of the rare examples of truly resonant architectures supporting disk-induced planet migration. The bright host star, $V \approx$ 11.9\,mag, the relatively short orbital period ($P_{\rm b}$ =  76.80$_{-0.06}^{+0.06}$\,d) and pronounced eccentricity (e =0.448$_{-0.029}^{+0.028}$) make the transiting planet a valuable target for atmospheric investigation with the {\em James Webb Space Telescope} (JWST) and ground-based extremely-large telescopes.

\end{abstract}

\keywords{methods: observational --
                techniques: radial velocities --
                techniques: photometric  -- planets and satellites: detection  -- planets and satellites: formation -- planets and satellites: dynamical evolution and stability
               }

%(No transits in Sector 34)

\section{Introduction}

The number of known exoplanets discovered by different surveys up to 2022 December is more than 5200, including around 850 multiple-planet systems\footnote{\url{http://exoplanet.eu/catalog/}}. 
%More than 2200 of these have a precise radius and mass determination.
The majority of these exoplanets have been detected using the transit technique, primarily thanks to the highly successful NASA \textit{Kepler} space telescope \citep{Borucki2010}, and the 
ongoing {\it Transiting Exoplanet Survey Satellite} \citep[$\tess$;][]{Ricker2015} mission.
However, the transit signals alone do not reveal the planetary mass and orbital eccentricity. A better picture of the distribution of the planetary radii, dynamical masses, bulk densities, and orbital geometry is fundamentally important for studying exoplanet composition, evolution, and overall formation. 
Therefore, validating and characterizing orbital and physical parameters of transiting planets with precise radial velocity (RV) data is fundamentally important to link observations with theory.
 
The $\tess$ survey has revealed many well-characterized exoplanet systems for which both mass and radius have been determined observationally. Indeed, the relatively small $\tess$ telescope is the main instrument for the discovery of transiting exoplanets around relatively bright stars, which allows them to be confirmed with ground-based precise RV measurements. 
A major role in the global efforts to characterize $\tess$ exoplanets is played by the {\bf W}arm g{\bf I}a{\bf N}ts with t{\bf E}ss (WINE) survey, which aims for the validation of warm (with periods ranging between $10\,\mathrm{days} < P < 300\,\mathrm{days}$) Jovian gas-giants by inspecting the $\tess$ Full Frame Images (FFI) data. 
Prospective targets are followed by an extensive RV monitoring with the FEROS\footnote{The Fiber-fed Extended Range Optical Spectrograph \citep[FEROS;][]{Kaufer1999} } and the HARPS\footnote{High Accuracy Radial velocity Planet Searcher at the ESO La Silla 3.6m telescope \citep[HARPS;][]{Mayor2003}} precise RV spectographs. The WINE survey has been highly successful, having detected and characterized many giant planets \citep[see e.g.][among many more]{Espinoza2020,jordan:2020,brahm:2020,Schlecker2020,Trifonov2021a}.\looseness=-5

 \begin{figure*}[tp]
    \centering
%$\begin{array}{cc} 
    \includegraphics[width=8.9cm]{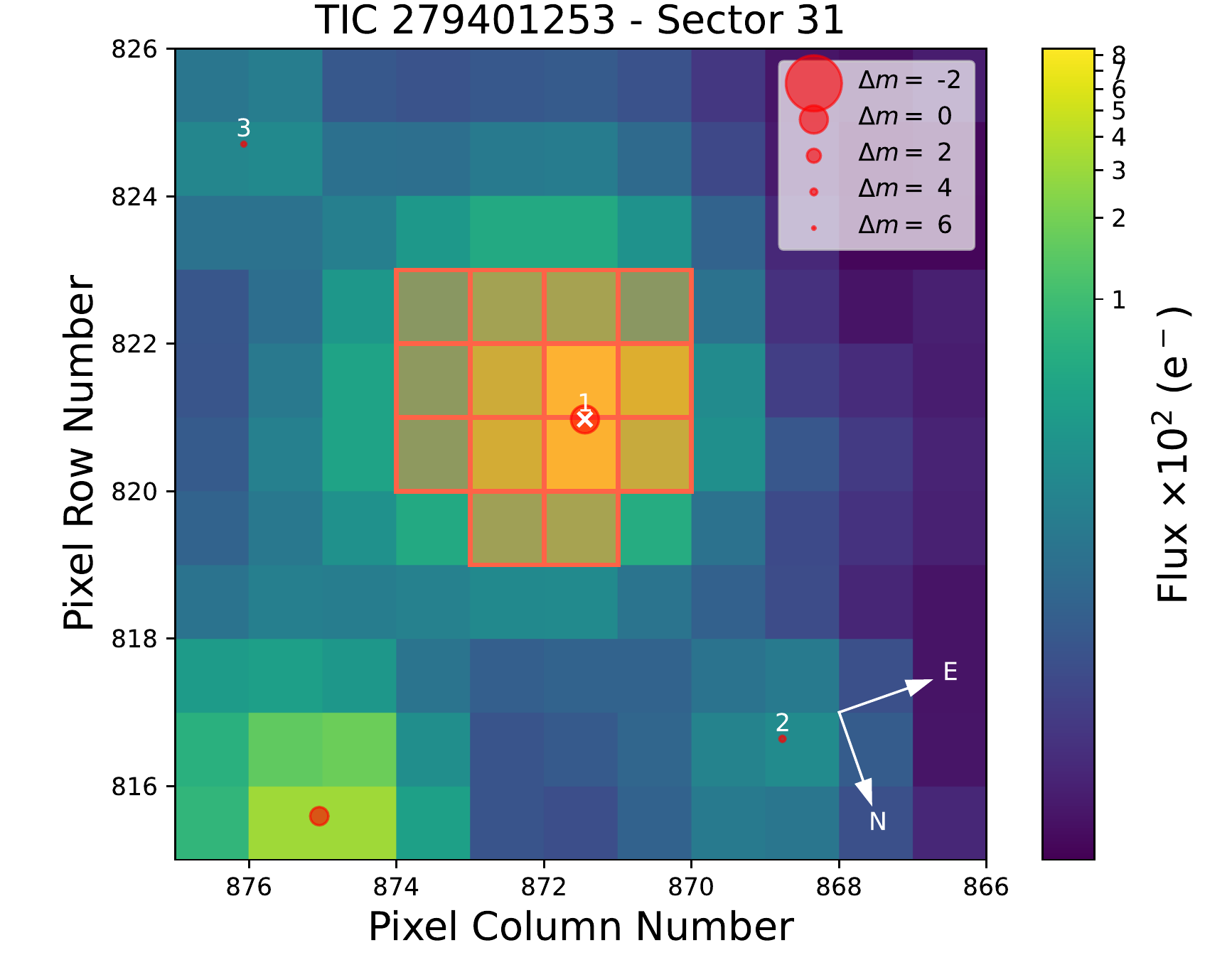} 
    \includegraphics[width=8.9cm]{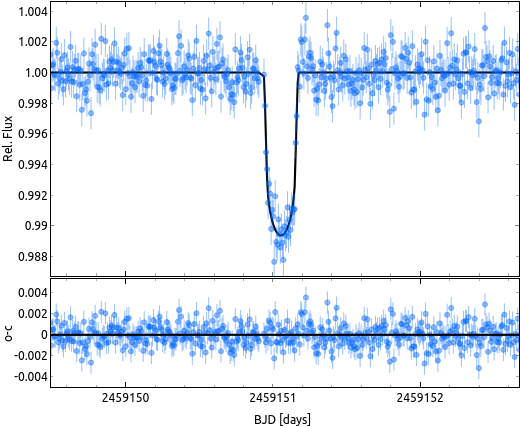}\put(-225,195){\normalsize TESS 10-min FFI} \\

    \caption{The left panel shows the TPF image of TIC\,279401253 for $\tess$ Sector 31. The red pixels indicate the combined aperture used to compute the photometry. The {\em Gaia} targets are shown with red circles, whose size is coded by their G magnitude. The right panel shows the {\sc tesseract} 10-minute FFI light curve of TIC\,279401253 around the single-transit event recorded in Sector\,31. The model curve shows a transit model to the data, and the small bottom panel shows the residuals of the fit.
    }
    \label{tpf}
\end{figure*}

In this paper, we report the discovery of a warm massive exoplanet pair around a G-dwarf star, which has been uncovered within the WINE-$\tess$ follow-up survey, based on RVs obtained with HARPS and FEROS. 
We present the TIC\,279401253 two-planet system, which exhibits a significant single transiting event detected in $\tess$,
consistent with a Jovian-sized planet. However, the acquired RVs of the TIC\,279401253 system undoubtedly reveal a strongly interacting warm giant-mass pair of planets locked in a 2:1 mean motion resonance (MMR) commensurability. The detection and observational characterization of warm Jovian planets in 2:1 MMR is still a rare event, despite its importance for constraining planet migration.
\looseness=-5

The observational data is presented in \autoref{sec2}. We use this  data to detect and characterize the warm pair of planets that orbit around TIC\,279401253. We present our estimates of the stellar parameters of TIC\,279401253 in \autoref{sec3}, along with the planetary orbital analysis performed jointly with the acquired Doppler data and $\tess$ photometry.
In \autoref{sec3}, we also provide our results from an analysis on the dynamical architecture and long-term stability of the TIC\,279401253 system. Finally, in \autoref{sec4}, we present a brief summary and our conclusions.

\section{Data}
\label{sec2}

\subsection{TESS}
\label{Sec2.1}

TIC\,279401253 was visited by $\tess$ during Sectors 4 and 31. Our team identified a single-transit event in the light curves extracted from the $\tess$ 10-minute Full-frame Images (FFIs) of Sector 31, using the {\sc tesseract}\footnote{\url{https://github.com/astrofelipe/tesseract}} pipeline (F. I. Rojas et al. 2023 in preparation). 
The left panel of \autoref{tpf} shows the target pixel file (TPF) image of TIC\,279401253\, constructed from the $\tess$ FFI image frames and {\em Gaia} DR3 data \citep{Gaia_Collaboration_2021}. The right panel of 
\autoref{tpf} shows the detrended {\sc tesseract} FFI light curve centered around the transit event, together with a transit model fit to the data. We did not identify bright contaminators in the FFI aperture (red continuous contour); thus we concluded that the transit signals are indeed coming from TIC\,279401253 and not from neighbouring stars.\looseness=-5 
 
The 2-minute cadence light-curves are retrieved from the Mikulski Archive for Space Telescopes\footnote{\url{https://mast.stsci.edu/portal/Mashup/Clients/Mast/Portal.html}}. Simple aperture photometry (SAP) and systematics-corrected Pre-search Data Conditioning photometry \citep[PDC,][]{Smith2012, Stumpe12} are provided by The Science Processing Operations Center \citep[SPOC;][]{SPOC} . 
The PDCSAP light curves are corrected for contamination from nearby stars and instrumental systematics originating from pointing drifts, focus changes, etc., and for contamination from nearby stars.

\subsection{RV Data}

After the detection of the $\tess$ transit, we initiated a 
Doppler follow-up campaign by obtaining precise RV data with the FEROS and HARPS spectrographs. We obtained 19 spectra of TIC\,279401253 with FEROS between 2021 February and 2022 October , and 14 spectra with HARPS  between 2021 September and 2022 October . 
With both instruments, we recorded stellar spectra in conjunction 
with a simultaneous ThAr lamp used for wavelength calibration. The exposure times were set to 1800 seconds, yielding an average signal-to-noise ratio per spectral resolution element of 76 for FEROS, and 34 for HARPS, respectively. 

 \begin{figure}[t]
\includegraphics[width=8.5cm]{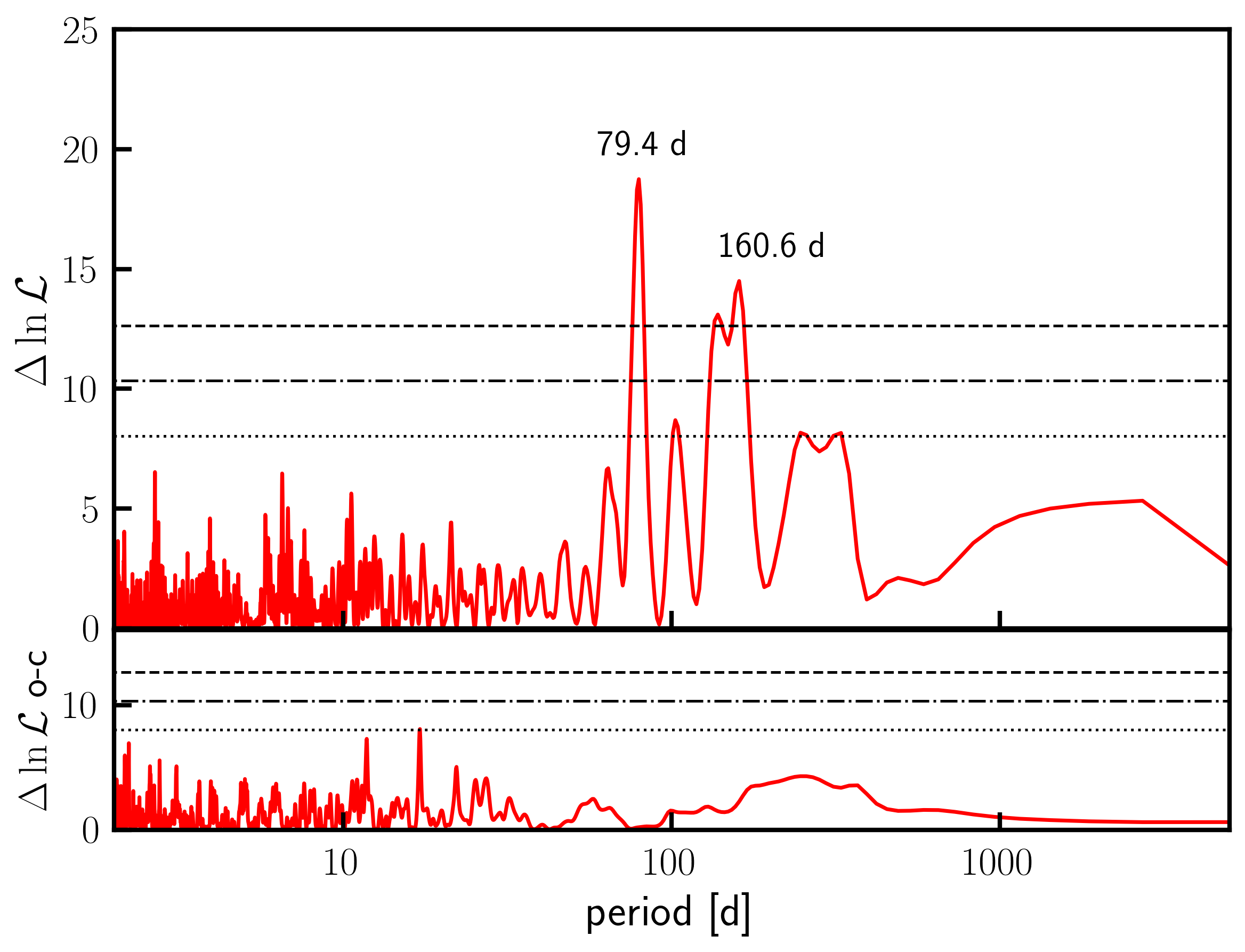}
\centering
\caption{\label{fig:MLP}
The MLP power spectrum of the combined FEROS and HARPS Doppler measurements of TIC\,279401253.  Horizontal lines reflect the $\Delta lnL$ values, which correspond to  0.1\%, 1\% and 10\% false alarm probabilities (from top to bottom). Two significant periods are detected: one near 79.4 days and the other near 160.6 days.The bottom panel shows the residuals of our two-planet best fit model (see \autoref{NS_params}), no significant peaks are observed.
}
\end{figure}

The FEROS spectra were reduced, extracted, and analysed with the \texttt{ceres} pipeline \citep{ceres}. For the HARPS spectra, we retrieved precise RV measurements derived by the ESO-DRS pipeline. Both \texttt{ceres} and ESO-DRS use a spectrum cross-correlation function (CCF) method with a weighted binary mask \citep{Pepe2002}.
With \texttt{ceres}, we measure FEROS RVs and  bisector span measurements with a mean uncertainty of $\hat\sigma_{\rm FEROS} = $ 8.2 m\,s$^{-1}$.
The DRS pipeline also provides the CCF's full-width half-maximum (FWHM) and the Bisector Inverse Slope span (BIS-span) measurements, which are valuable stellar activity indicators \citep[][]{Queloz2001}. 
The mean RV uncertainty of ESO-DRS is $\hat\sigma_{\rm HARPS}$ = 2.9  m\,s$^{-1}$. 
When combined in common mean RV offset, the Doppler velocities show very large end-to-end periodic RV amplitude of $\sim$ 880 m\,s$^{-1}$, suggesting suggesting a massive substellar companion (or companions).. We did not detect any significant periodicity in the FEROS and HARPS activity data.
The obtained FEROS RVs are presented in \autoref{table:FEROS_RVs}, and the precise HARPS-DRS RVs and activity index data are tabulated in \ref{table:HARPS_RVs}.

\begin{table}[tp]
% \begin{adjustwidth}{-4.0cm}{} 

\caption{Stellar parameters of TIC\,279401253 and their 1$\sigma$ uncertainties derived using 
ZASPE analyses of HARPS spectra, {\em Gaia DR3} parallax, broadband photometry, and PARSEC models.}
\label{table:phys_param}    

%(((0,849+0,024)×0,05)+ ((0,849−0,033)×0,05))÷2

\centering

\begin{tabular}{ p{6.0cm}  r r}     % 3 columns 
\hline\hline  \noalign{\vskip 0.5mm}        
  Parameter   &    \\  
\hline    \noalign{\vskip 0.5mm}                   
 %  Spectral type                            & G3V          & [1] \\ 
   Distance$^1$  (pc)                       & 287.1$_{-1.9}^{+1.9}$    \\   \noalign{\vskip 0.9mm}
   Mass    ($M_{\odot}$)                    & 1.13$_{-0.03}^{+0.02}$   \\  \noalign{\vskip 0.9mm}
   Radius    ($R_{\odot}$)                  & 1.06$_{-0.01}^{+0.01}$      \\ \noalign{\vskip 0.9mm}
   Luminosity    ($L{_\odot}$)              & 1.25$_{-0.04}^{+0.05}$    \\ \noalign{\vskip 0.9mm}
   Age (Gyr)                                & 1.2$_{-0.8}^{+1.0}$     \\   \noalign{\vskip 0.9mm}
   A$_V$   (mag)                            & 0.09$_{-0.05}^{+0.06}$     \\ \noalign{\vskip 0.9mm}
   $T_{\mathrm{eff}}$~(K)                   & 5951$\pm$80      \\ \noalign{\vskip 0.9mm}
   $\log g~[\mathrm{cm\cdot s}^{-2}]$       & 4.438$\pm$0.015      \\    \noalign{\vskip 0.9mm}
   {}[Fe/H]                                 & 0.20$\pm$0.05      \\ \noalign{\vskip 0.9mm}
   $v\cdot\sin(i)$ (km\,s$^{-1}$)           & 5.0$\pm$0.5     \\         \noalign{\vskip 0.9mm}                                  
 
\hline\hline \noalign{\vskip 0.5mm}   
 
\end{tabular}
 
% \end{minipage}}
% \end{adjustwidth}
 
\tablecomments{
1 -- \citet{Gaia_Collaboration_2021}. 
%  %The values in parentheses are "floor" (i.e., more realistic, minimum) uncertainties predicted by \citet{Tayar2020} and adopted in our work.
}

\end{table}

\section{Analysis and results}
\label{sec3}

\begin{figure*}
    \centering
    \includegraphics[width=18cm]{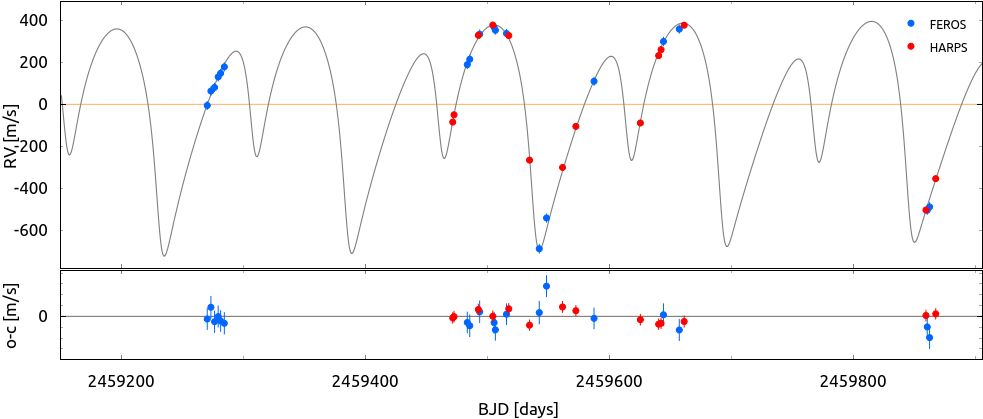} \\ \vspace{0.2cm}
        \includegraphics[width=5.9cm]{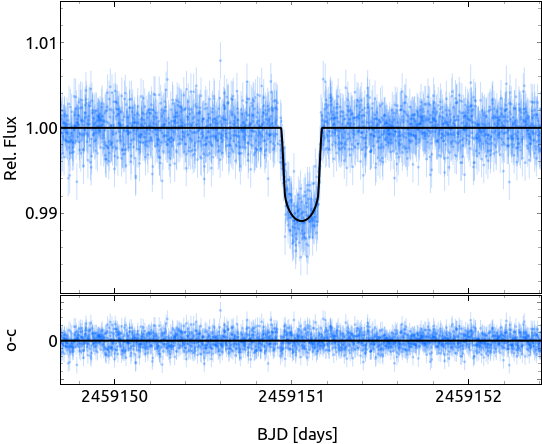} \put(-140,60){\scriptsize TIC\,279401253\,b} \put(-145,120){\scriptsize TESS 2-min PDCSAP}
    \includegraphics[width=5.9cm]{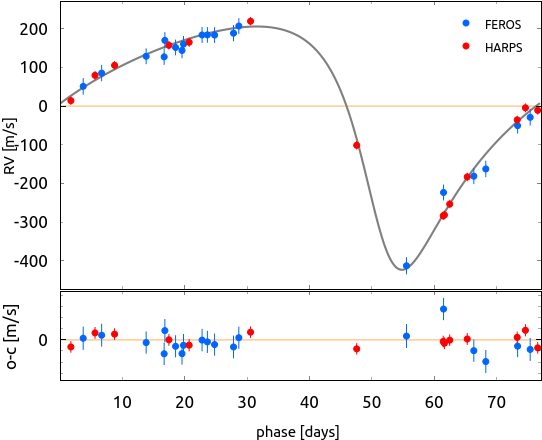} \put(-140,60){\scriptsize TIC\,279401253\,b}
    \includegraphics[width=5.9cm]{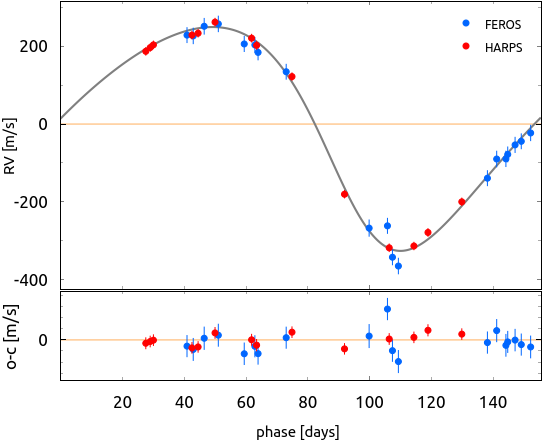} \put(-140,60){\scriptsize TIC\,279401253\,c} \\

    \caption{The top main panel shows the Doppler measurements from FEROS (blue) and  HARPS (red) and the best two-planet model for these combined data. The left bottom panel shows the PDCSAP 2-min $\tess$ light curve of TIC\,279401253 around the single-transit event recorded in Sector\,31. The model curve shows the transit counterpart of the best-fit model applied jointly to the Doppler data obtained with FEROS and HARPS. The middle and right bottom panels show a phase-folded representation of the two planetary signals after the RV signal of the other companion was subtracted. The respective residuals are shown under each panel, accordingly.} 
\label{fig:RV_curve} 
\end{figure*}

%The data uncertainties include the RV jitter.

\subsection{Stellar parameters}
\label{sec3.1}

TIC\,279401253 is a G-type star visible in the Southern Hemisphere. 
The star is at a distance of about 287.1$_{-1.9}^{+1.9}$ pc from the Sun and has an apparent magnitude of V = $11.9$ mag.
The atmospheric and physical parameters were obtained using three coadded HARPS spectra and the {\tt ZASPE} code \citep[][]{zaspe}. For TIC\,279401253 we obtain an effective temperature of $T_{\rm eff} = 5951 \pm 80$ K, a metallicity of $[{\text{Fe/H}}] =  +0.20\pm 0.05$, and a projected rotational velocity of $v\text{sin} i = 5.0 \pm 0.5$ km\,sec$^{-1}$.
Further, we derive stellar physical parameters using the PARSEC stellar isochrones \citep{parsec}, following the recipe in \citet{Brahm2018} in conjunction with the {\em Gaia}  parallaxes, and the public broadband photometry (G,G$_{BP}$, G$_{RP}$, J, H, K). We derive a stellar mass of M$_\star = 1.13^{+0.02}_{-0.03}$ M$_{\odot}$ and a stellar radius of R$_\star = 1.06\pm 0.01$ R$_{\odot}$. We list the remaining atmospheric and physical parameters in \autoref{table:phys_param}.

\subsection{Orbital analysis}

For the transit, RV, and joint RV-transit analyses, we use the {\tt Exo-Striker} exoplanet toolbox\footnote{\url{https://github.com/3fon3fonov/exostriker}} \citep{Trifonov2019_es}. {\tt Exo-Striker} allows for the use of multiple-Keplerian or self-consistent N-body dynamical models.  During the course of our analysis, though, we concluded that a single-transit event and the relatively sparse RV data led to strong degeneracies in the N-body model. Therefore, in this work, we only make use of the Keplerian model. The parameters in our transit model are: period $P$, eccentricity $e$, argument of periastron $\omega$, inclination $i$, time of inferior transit conjunction $t_{0}$, and the planetary semi-major axis $a/R_\star$, and radius $r/R_\star$ (in relative stellar units), respectively. Additionally, we use quadratic limb darkening parameters for modeling the light curve.
The parameters in our RV model are: RV semiamplitude $K$, orbital period~$P$, eccentricity $e$, argument of periastron $\omega$, and mean anomaly $M_0$. All these parameters are valid for BJD = 2459151.0, which was deliberately chosen slightly before the epoch of the $\tess$ mid-transit event. Additional fitting parameters in our RV modeling of TIC\,279401253 were the FEROS and HARPS RV data offsets and variance \citep[i.e., RV jitter,][]{Baluev2009}. 
 
For deriving the best-fit orbital parameters, we adopted a maximum likelihood estimator (MLE) scheme, which optimizes the parameters via the Nelder-Mead simplex algorithm \citep[][]{NelderMead}, followed by a Levenberg-Marquardt (LM) algorithm \citep[][]{Press}. These best-fit estimates and the LM covariance matrix confidence intervals serve as prior knowledge for our more detailed Bayesian posterior analysis of the orbital parameters. The latter was constructed using the {\tt dynesty} sampler \citep{Speagle2020}, which utilizes a nested sampling scheme \citep[NS, ][]{Skilling2004}. 
 
\subsubsection{$\tess$ analysis}
\label{4.1.1}

Since the $\tess$ light curve contains of only a single-transit event, no meaningful period estimate could be derived. However, the transit shape depends on the orbital inclination, the planet radius, $r_p/R_\star$, and the ratio of the semi-major axis of the planetary orbit to the stellar radius, $a_p/R_\star$. The latter is related to Kepler's third law, from which we could obtain a crude estimate of the orbital period \citep[e.g.][]{sandford:2019}. Assuming a circular orbit, we fit the $\tess$ light curve with a Keplerian model, and we extract a mid-transit time t$_0$ = 2459151.054$_{-0.001}^{+0.001}$ BJD, an inclination i = 89.8$_{-0.1}^{+0.1}$ deg, a companion radius R$_b$ = 1.00$_{-0.03}^{+0.03}$ R$_{\rm Jup}$, and an orbital period P = 82.9$_{-22.2}^{+23.1}$ days. Based on these estimates, we initiated an RV follow-up within the WINE survey to confirm and characterize the planetary companion.

\begin{table*}[ht]

\centering   
\caption{{Nested sampling priors, posteriors, and the optimum $-\ln\mathcal{L}$ orbital parameters of the two-planet system TIC\,279401253 derived by joint modeling of $\tess$ data, and RVs from FEROS and HARPS.}}
\label{NS_params}

 \begin{adjustwidth}{-2.2cm}{}
 \resizebox{0.85\textheight}{!}
 {\begin{minipage}{1.1\textwidth}

\begin{tabular}{lrrrrrrrrrrrr}     % 2 columns

\hline\hline  \noalign{\vskip 0.7mm}

%\makebox[0.1\textwidth][l]{\hspace{60 mm}$\sim$2:1 MMR fit      \hspace{60 mm} $\sim$3:1 MMR fit  \hspace{1.5 mm} } \\
%\cline{2-5}\cline{7-10}\noalign{\vskip 0.9mm}   
\makebox[0.1\textwidth][l]{\hspace{50 mm} Median and $1\sigma$  \hspace{20 mm} Max. $-\ln\mathcal{L}$     \hspace{30 mm} Adopted priors  \hspace{10 mm} \hspace{1.5 mm} } \\
\cline{1-9}\noalign{\vskip 0.7mm}

Parameter &\hspace{10.0 mm} Planet b & Planet c &  & Planet b & Planet c  & & \hspace{10.0 mm}Planet b & Planet c  \\
%   \hline \noalign{\vskip 0.7mm}
\cline{1-9}\noalign{\vskip 0.7mm}

%K$_b$ [m/s] = 301.9947051698501 - 20.98356436339992 + 22.618855906036174
%P$_b$ [d] = 76.80037501544221 - 0.06555595880617204 + 0.0641791449129272
%e$_b$ = 0.44852051637142804 - 0.029041034308471658 + 0.02787219961731341
%$\omega_b$ [deg] = 137.58322337638197 - 6.700477693689834 + 6.1236151506007275
%K$_c$ [m/s] = 286.9154805256775 - 5.669402427954935 + 5.89582458417442
%P$_c$ [d] = 155.30805402658612 - 0.6964908853801433 + 0.6969274922049635
%e$_c$ = 0.25426987275343055 - 0.04171002297295784 + 0.036344306661769354
%$\omega_c$ [deg] = 120.94483380025466 - 8.071979123090713 + 7.894593336215593

%Median and their 1 sigma errors

%i$_b$ [deg] = 89.59492146742645 - 0.2592268004387819 + 0.2756522151032357
%t0 $b$ [d] = 2459151.0552331647 - 0.0011614696122705936 + 0.0011350787244737148
%a/$R_\star$ $b$ = 71.90100881256613 - 4.462743622319721 + 5.192482411088022
%R/$R_\star$ $b$ = 0.09438658544116137 - 0.0036877955206489482 + 0.0032962494278445525
%t0 $c$ [d] = 2459216.2197749796 - 2.2732872213236988 + 2.3137567550875247
%transit off$_1$ [ppm] = 174.25041533257814 - 31.978764397391956 + 32.334911628296055
%transit jitt$_1$ [ppm] = 29.912547277325043 - 15.782662383559929 + 35.14036835342033
%ld-quad-1$_1$ [ppm] = 0.3563607896076184 - 0.18910809795547148 + 0.18655550997813208
%ld-quad-2$_1$ [ppm] = 0.5193825728259304 - 0.3378939146854375 + 0.31877519622008443 

$K$  [m\,s$^{-1}$]            &  302.0$_{-21.0}^{+22.6}$ & 286.9$_{-5.7}^{+5.9}$ &  
                              &  314.6& 287.5 &
                              &  $\mathcal{U}$(200.0,400.0) & $\mathcal{U}$(200.0,400.0)  &  \\ \noalign{\vskip 0.9mm}

$P$  [day]                    & 76.80$_{-0.06}^{+0.06}$ & 155.3$_{-0.7}^{+0.7}$ & 
                              & 76.82 & 155.25 &  
                              & $\mathcal{U}$(74.0,80.0) & $\mathcal{U}$(150.0,160.0)   &  \\ \noalign{\vskip 0.9mm}
                              
$e$                           & 0.448$_{-0.028}^{+0.029}$ & 0.254$_{-0.036}^{+0.042}$ &
                              & 0.454 &  0.215 &  
                              & $\mathcal{N}$(0.0,0.2) &  $\mathcal{N}$(0.0,0.2)  &  \\ \noalign{\vskip 0.9mm}

$\omega$  [deg]               & 137.6$_{-6.7}^{+6.1}$  & 120.9$_{-8.1}^{+7.9}$ &
                              & 139.5  & 128.2&   
                              & $\mathcal{U}$(0.0,360.0) &  $\mathcal{U}$(0.0,360.0) &  \\ \noalign{\vskip 0.9mm}
                              
$t_{\rm 0}$ - 2459000  [BJD]   & 151.055$_{-0.001}^{+0.001}$ &   $\dots$&
                              & 151.055   & $\dots$&  
                              & $\mathcal{U}$(151.0,151.1) & $\dots$    &  \\ \noalign{\vskip 0.9mm}
 
$M_{\rm 0}$  [deg]            & $\dots$   & 190.7$_{-10.2}^{+9.6}$ &
                              & $\dots$   & 182.4 &  
                              & $\dots$  &  $\mathcal{U}$(0.0,360.0)  &  \\ \noalign{\vskip 0.9mm}

%$\lambda$  [deg]          &  107.0$_{-0.8}^{+0.8}$  & 93.4$_{-1.0}^{+0.9}$ & 
%                              & 107.1  & 93.1 &    
%                              &  (derived) &   (derived)  &  \\ \noalign{\vskip 0.9mm}
 
$i$          [deg]            & 89.59$_{-0.25}^{+0.27}$  & 90.0 &
                              & 89.56  & 90.0 & 
                              & $\mathcal{N}$(90.0,0.5) &  (fixed) &  \\ \noalign{\vskip 0.9mm}  
                              %& $\mathcal{U}$(85.0,95.0) &  $\mathcal{U}$(85.0,95.0)  &  \\ \noalign{\vskip 0.9mm}  
$a_p$/$R_\star$                  & 71.9$_{-4.4}^{+5.2}$  &  \dots &
                              & 75.2  & \dots & 
                              & $\mathcal{U}$(65.00,85.00) &  \dots &  \\ \noalign{\vskip 0.9mm}  
                              
$r_p$/$R_\star$                  & 0.094$_{-0.004}^{+0.003}$  &  \dots &
                              & 0.097  & \dots & 
                              & $\mathcal{U}$(0.08,0.12) &  \dots &  \\ \noalign{\vskip 0.9mm}                              
%a/$R_\star$ $b$ = 71.90100881256613 - 4.462743622319721 + 5.192482411088022
%R/$R_\star$ $b$ = 0.09438658544116137 - 0.0036877955206489482 + 0.0032962494278445525

%$\Omega$     [deg]            & 0.0   & 2.0$_{-1.2}^{+ 1.9}$ &
%                              & 0.0  & 1.9&  
%                              & (fixed) &  $\mathcal{N}$(0.0,15.0)  &  \\ \noalign{\vskip 0.9mm}                               

% $\Delta i$  [deg]             & 2.0$_{-1.2}^{+1.9}$  & $\dots$ &  
%                               &  1.9 & $\dots$ &     
%                               &  (derived) &   $\dots$  &  \\ \noalign{\vskip 0.9mm}

$a$  [au]                     &  0.369$_{-0.003}^{+0.003}$  & 0.591$_{-0.006}^{+0.005}$  &  
                              &  0.369 & 0.591&     
                              &  (derived) &   (derived)  &  \\ \noalign{\vskip 0.9mm}

$m$  [$M_{\rm jup}$]          & 6.14$_{-0.42}^{+0.39}$  & 8.02$_{-0.18}^{+0.18}$ & 
                              & 6.38 & 8.12 &     
                              &  (derived) &   (derived)  &  \\ \noalign{\vskip 0.9mm}

$R$  [$R_{\rm jup}$]          & 1.00$_{-0.04}^{+0.04}$  & $\dots$ & 
                              & 1.03  & $\dots$ &     
                              &  (derived) &   $\dots$  &  \\ \noalign{\vskip 0.9mm}    

%m$_b$ [M$_{\rm Jup}$] = 6.1404635932014155 - 0.3914888455550436 + 0.4190962854670994
%m $\sin i_c$ [M$_{\rm Jup}$] = 8.022825849867388 - 0.17935669769176243 + 0.17961027302437316
%a$_b$ [au] = 0.368932923923602 - 0.003308373157084321 + 0.0032474255226619952
%a$_c$ [au] = 0.5913025095868877 - 0.0055545045842753105 + 0.005472195338530161
%R$_b$ [R$_{\rm Jup}$] = 0.9959119596712281 - 0.03996636212393678 + 0.03569369057234251
                              
%\accepted{\cline{1-5}\cline{7-10}\noalign{\vskip 0.7mm}  
% $\rho$ [g\,cm$^{-3}$]       & 0.174$_{-0.015}^{+0.016}$ & 1.014$_{-0.076}^{+0.084}$ & 
%                              & 0.174 & 1.014 &     
 %                             &  (derived) &   (derived)  &  \\ \noalign{\vskip 0.9mm}

RV$_{\rm off.}$ FEROS  [m\,s$^{-1}$]                &     14576.1$_{-6.5}^{+6.5}$ &   $\dots$  &     &     14579.8 &  $\dots$    &      & $\mathcal{U}$(14000.0,15000.0)&   $\dots$ \\ \noalign{\vskip 0.9mm}
RV$_{\rm off.}$ HARPS  [m\,s$^{-1}$]               &      14595.3$_{-5.6}^{+5.9}$  &   $\dots$ &   &    14594.1     &  $\dots$   & &$\mathcal{U}$(14000.0,15000.0))&  $\dots$ \\ \noalign{\vskip 0.9mm}

RV$_{\rm jit.}$ FEROS  [m\,s$^{-1}$]               &      19.9$_{-3.8}^{+4.9}$  & $\dots$  &   & 17.6&  $\dots$    &         & $\mathcal{J}$(0.0,50.0)&   $\dots$  \\ \noalign{\vskip 0.9mm}
RV$_{\rm jit.}$ HARPS  [m\,s$^{-1}$]                &    9.9$_{-2.5}^{+3.9}$  &  $\dots$ &    & 9.2 &    $\dots$   &          & $\mathcal{J}$(0.0,50.0)&  $\dots$    \\ \noalign{\vskip 0.9mm}

Tran.$_{\rm off.}$ TESS--S31 [ppm]               &     174$_{-32}^{+32}$  & $\dots$  &   & 173&  $\dots$    &         & $\mathcal{U}$(-1000.0,1000.0)&   $\dots$  \\ \noalign{\vskip 0.9mm}
Tran.$_{\rm jit.}$ TESS--S31 [ppm]                &    30$_{-16}^{+35}$  &  $\dots$ &    & 37 &    $\dots$   &          & $\mathcal{J}$(0.0,100)&  $\dots$    \\ \noalign{\vskip 0.9mm}

LD-quad.$_{\rm 1}$ TESS--S31             &     0.36$_{-0.19}^{+0.19}$  & $\dots$  &   & 0.44&  $\dots$    &         & $\mathcal{U}$(0.0,1.0)&   $\dots$  \\ \noalign{\vskip 0.9mm}
LD-quad.$_{\rm 2}$ TESS--S31                 &   0.52$_{-0.34}^{+0.32}$  &  $\dots$ &    & 0.65 &    $\dots$   &          & $\mathcal{U}$(0.0,1.0)&    $\dots$    \\ \noalign{\vskip 0.9mm}

%RV off$_{\rm FEROS}$ [m/s] = 14576.056516653918 - 6.4614698595541995 + 6.483803054761665
%RV off$_{\rm HARPS}$ [m/s] = 14595.28270930708 - 5.595870952054611 + 5.8829381948398805
%RV jitt$_{\rm FEROS}$ [m/s] = 19.94474700831363 - 3.836273334338099 + 4.933609641643564
%RV jitt$_{\rm HARPS}$ [m/s] = 9.882098820816614 - 2.5605565109616544 + 3.931535860873126
%transit off$_1$ [ppm] = 174.25041533257814 - 31.978764397391956 + 32.334911628296055
%transit jitt$_1$ [ppm] = 29.912547277325043 - 15.782662383559929 + 35.14036835342033
%ld-quad-1$_1$ [ppm] = 0.3563607896076184 - 0.18910809795547148 + 0.18655550997813208
%ld-quad-2$_1$ [ppm] = 0.5193825728259304 - 0.3378939146854375 + 0.31877519622008443 

\\
\hline \noalign{\vskip 0.7mm}

\end{tabular}

\end{minipage}}
\end{adjustwidth}
\tablecomments{The orbital elements are in the Jacobi frame and are valid for epoch BJD = 2459151.0. The adopted priors are listed in the right-most columns and their meanings are $\mathcal{U}$ -- Uniform, $\mathcal{N}$ -- Gaussian, and $\mathcal{J}$ -- Jeffrey's (log-uniform) priors. The derived planetary posterior parameters of $a$, $m$, and $R$ are calculated taking into account the stellar parameter uncertainties.
 }
\end{table*}

\subsubsection{RV-only analysis}
\label{3.2.2}

The precise Doppler data of TIC\,279401253 exhibited strong periodicity, which could be attributed to a planetary companion. We performed a period search analysis by computing a maximum likelihood periodogram \citep[MLP;][]{Baluev2008} to the combined HARPS and FEROS data set. The MLP fits a sine curve to the RV data for a given frequency grid and optimizes the semiamplitude, phase, RV offset, and  RV jitter parameters of FEROS and HAPRS. The resulting $\Delta \ln\mathcal{L}$ power spectrum is shown on \autoref{fig:MLP}. We detected two significant periods, one near 79.4 days and the other near 160.6 days. 

A single Keplerian with a base period of 79.4\,d did not lead to an adequate fit to the RV data. An MLP to the residuals confirmed the $\sim$ 160\,d significant period. The combined RV data is consistent with two strong periodic signals in the 2:1 period ratio commensurability.
We found that a two-planet Keplerian model presented an excellent RV solution with no significant periodicities left in the residuals. Our MLE best-fit suggests planetary periods of $P_{\rm b}$ = 77.21$\pm$0.07\,d, and $P_{\rm c}$ = 154.5$\pm$0.4\,d, RV semiamplitudes of  $K_{\rm b}$ = 340.7$\pm$12.1\,m\,s$^{-1}$, and $K_{\rm c}$ = 265.6$\pm$4.1\,m\,s$^{-1}$, and eccentricities of $e_{\rm b}$ = 0.46$\pm$0.02\,d, and $e_{\rm c}$ = 0.18$\pm$0.04\,d, for the inner and the outer planet, respectively. We derived minimum planetary masses of $m_b\sin(i)$ $\sim$ 6.9$M_{\rm jup}$, and  $m_c\sin(i)$ $\sim$ 7.5 $M_{\rm jup}$, respectively.

\subsubsection{RV-TESS joint analysis}
\label{sec3.2.3}

We use the individual transit and RV parameters' MLE estimates to define 
our prior ranges. Then we run a joint MLE fit and NS global parameter posterior analysis. The single-transit event, together with the evidence of two similar Jovian-mass planets, poses the question, which planet actually transits? We tested both possibilities and could construct a consistent joint fit only when the $\tess$ transit event is related to the inner planet, TIC\,279401253\,b. Therefore, our work only discusses fits with TIC\,279401253\,b being the transiting planet.

\begin{figure*}
\centering
\includegraphics[width=5.9cm]{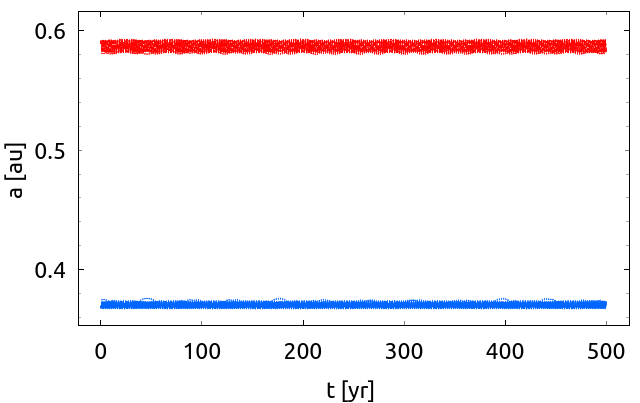}
\includegraphics[width=5.9cm]{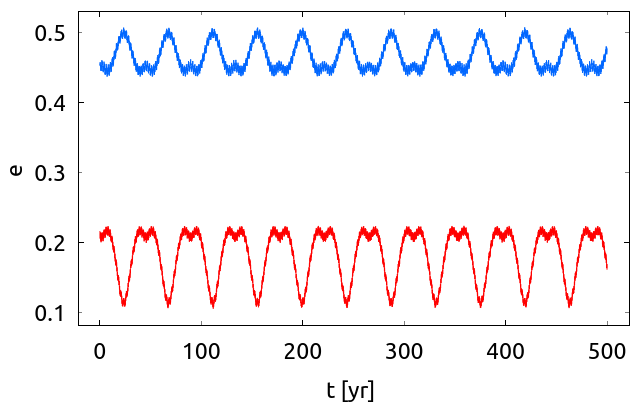} 
\includegraphics[width=5.9cm]{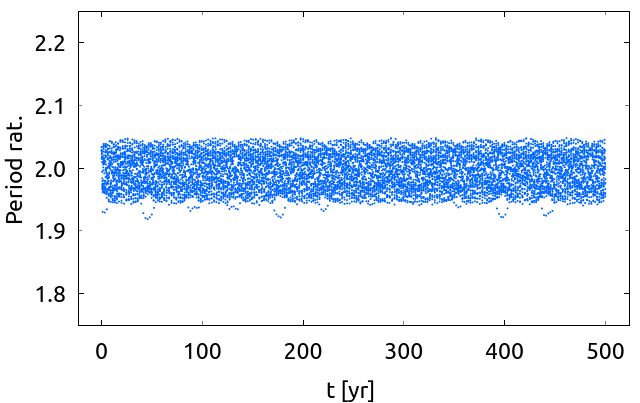}\\
\includegraphics[width=5.9cm]{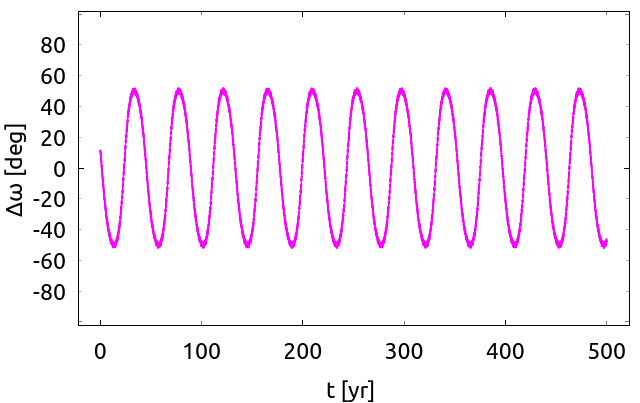}
\includegraphics[width=5.9cm]{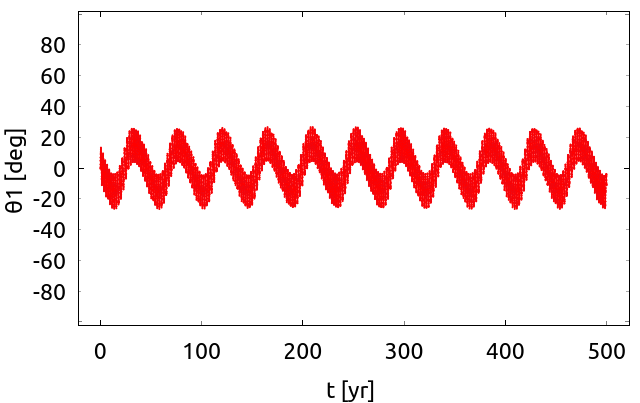} 
\includegraphics[width=5.9cm]{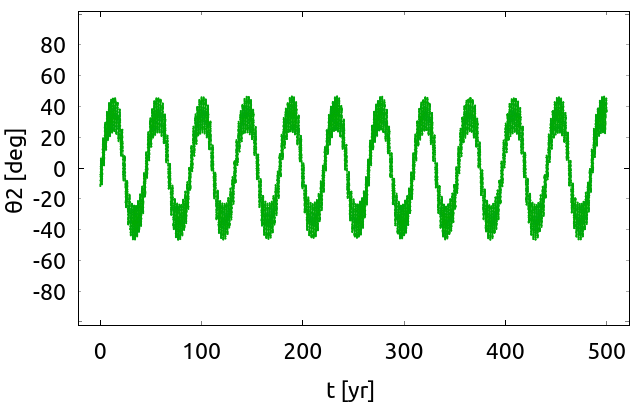}
\caption{Orbital evolution of the TIC\,279401253 system for a 500\,yr extent for the best fit solution. The top panels show the evolution of the semi-major axes $a_{\rm b,c}$, the eccentricities $e_{\rm b,c}$ (blue colour indicates the inner, red colour  the outer planet),
and planetary period ratio $P_{\rm rat.}$.  The bottom panels show the evolution of 
the apsidal alignment argument $\Delta\omega$ = $\omega_c$ - $\omega_b$ and the resonance angles $\theta_1$ and $\theta_2$, which librate around 0$^{\circ}$, i.e., in a 2:1 MMR. See text for details.}
\label{evol_plot} 
\end{figure*}

We performed an NS run, which allowed us to efficiently explore the parameter space of orbital elements and study the probability distribution of the posteriors. We ran 100 "live-points" per fitted parameter using the "Dynamic" NS scheme, focused on 100\% posterior convergence instead of log-evidence \citep[see,][for details]{Speagle2020}. The final adopted parameter priors, posteriors, and best-fit solution are listed in \autoref{NS_params}. The right panel of \autoref{tpf} shows the transit joint model counterpart to the $\tess$ data.
\autoref{fig:RV_curve} shows the RV data together with the best-fit Doppler joint model of TIC\,279401253 constrained by the single $\tess$ transit event. The middle and right bottom panels of \autoref{fig:RV_curve} show a phase-folded representation of the RV signals of TIC\,279401253\,b \& c, respectively. 
 There are no significant periods left in the 
 residuals of this fit, as can be seen from the bottom panel of \autoref{fig:MLP}.
 The final posterior probability distributions are shown in \autoref{Kep_cornerplot}. We note that our analysis is not strictly coplanar, as can be seen from table \autoref{NS_params}. While we fit the inclination of the transiting planet, the inclination of the outer planet remains fixed at 90 deg (and $\Delta\Omega$ = 0 deg). Therefore, our posterior distribution is consistent with a nearly coplanar edge-on system, which is a plausible outcome of the system architecture. The real mutual inclination, however, could be larger and is not possible to be revealed given the available data. Future additional transit and RV data in conjunction with photo-dynamical modeling could reveal the system architecture better.
Furthermore, RV measurements during the transit events could measure the magnitude of the Rossiter-Mclaughlin effect, providing a valuable insight into the spin-orbit alignment of the planet's orbit with respect to the stellar rotation, thus offering a powerful tool to study its formation and subsequent orbital evolution. However, a more precise constraint on the transit timing would be required to make these observations feasible.

Our final estimates for TIC\,279401253\,b \& c lead to planetary orbital periods of 
$P_b$ = 76.80$_{-0.06}^{+0.06}$ days, and 
$P_c$ = 155.3$_{-0.7}^{+0.7}$ days, 
eccentricities of  
$e_b$ = 0.448$_{-0.028}^{+0.029}$ and 
$e_c$ = 0.254$_{-0.036}^{+0.042}$.  We measured a dynamical mass of
$m_b$ = 6.14$_{-0.42}^{+0.39}$$M_{\rm jup}$, and  a minimum mass of
$m_c\sin(i)$ = 8.02$_{-0.18}^{+0.18}$~$M_{\rm jup}$, for the inner and outer planet, respectively.

%Based on the available data, we have no direct evidence that TIC\,279401253 c is a transiting planet. so we only assume a minimum mass, and therefore, an orbital inclination of 90 deg. 

\subsection{TIC\,279401253: A 2:1 MMR exoplanet system}
\label{sec3.3}

The companion periods, eccentricities, and masses in the TIC\,279401253 system point to a strongly interacting planet pair, which is likely involved in a 2:1 mean motion resonance (MMR). Therefore, we performed detailed numerical orbital evolution simulations to test this possibility and study the system's dynamical architecture. We use a custom version of the 
%Wisdom-Holman $N$-body algorithm \citep[][]{Wisdom1991} integrated in 
{\sc SyMBA} symplectic $N$-body algorithm \citep[][]{Duncan1998} integrated in the {\tt Exo-Striker} toolbox, which directly adopts and integrates the Jacobi orbital elements from the posterior orbital analysis.   We tested the stability of the TIC\,279401253 system up to 1\,Myr with a small time-step of 0.2\,d for 5000 randomly chosen samples from the joint transit+RV orbital parameter posteriors. 
For each integrated sample, we automatically monitored the evolution of the planetary semi-major axes, eccentricities, secular apsidal angle $\Delta\omega$ = $\omega_{\rm b}$ - $\omega_{\rm c}$, and first-order 2:1 MMR angles $\theta_1 =  \lambda_{\rm b} - 2\lambda_{\rm c} + \omega_{\rm b},~~\theta_2 =  \lambda_{\rm b} - 2\lambda_{\rm c} + \omega_{\rm c}$, where $\lambda_{\rm b,c} = M_{\rm b,c} + \omega_{\rm b,c}$ is the mean longitude of planet b and c, respectively \citep[see, e.g.,][]{Lee2004}.

We found that 73.4\% of the examined 5000 samples are stable for 1\,Myr, whereas 
the best-fit, for which we ran a longer simulation, is stable for 10\,Myr. We found that the stable configurations exhibit common dynamical behaviour with the system locked in the 2:1 MMR. 
%This is confirmed by the libratation of $\theta_1$ and $\theta_2$. 
\autoref{evol_plot} shows an example of a 500\,yr extent of the orbital evolution simulation started from the best-fit (i.e., maximum $-\ln\mathcal{L}$, see \autoref{NS_params}). 
We show the evolution of the mutual period ratio $P_{\rm rat.}$, and of the eccentricities $e_{\rm b}$ and $e_{\rm c}$. The TIC\,279401253 system is stable and osculates in the eccentricities and in the 2:1 period ratio. The apsidal alignment argument $\Delta\omega$ librates around  0$^{\circ}$ with semiamplitude of $\sim$ 65$^{\circ}$, whereas the characteristic 2:1 MMR angles, $\theta_1$ and $\theta_2$, librate around 0$^{\circ}$ with semiamplitudes respectively of $\sim$ 30$^{\circ}$ and $\sim$ 55$^{\circ}$. Therefore, the massive planetary pair is locked in a 2:1 MMR with a short secular timescale of the order of $\sim$ 45\,yr. 

We observed a similar dynamical picture when we studied the stable posterior probability distribution. We found that 90\% of the stable samples are locked in 2:1 MMR with $\Delta\omega$, $\theta_1$ and $\theta_2$ librating around 0 deg, with a median semiamplitudes of 62.3 deg, 31.7 deg, and 54.8 deg, respectively.

\subsection{Transit predictions}
\label{sec3.4}
As we discussed in \autoref{sec3.3}, the 2:1 MMR system is dynamically active for a short time. Therefore, we expect strong transit timing variations (TTVs), which could be predicted for future observations.
We extracted transit predictions from our dynamical analysis of the posterior distribution. We found that accurate TTV predictions are very challenging due to the strong ambiguity in eccentricity versus dynamical planetary mass %space, which is well known in the literature \citep[e.g.,][]{Lithwick2012}. 
space \citep[][]{Lithwick2012}. 
Due to the large multimodality of the TTVs, the error in transit predictions rapidly accumulates and is close to a few tens of days in 2023 and accumulates even more in future epochs. 
%\textbf{Moreover, the mutual even if nearly coplanar is contributing to the difficulty of the task.} 
We tried to construct a joint N-body model \citep[e.g.,][]{Trifonov2021b} from the available data and therefore predict future transits from fitting alone. However, the single-transit event and the relatively sparse RV data manifest in strong N-body model degeneracy in the posteriors, which is consistent with the TTV prediction from our stability analysis. 
 The TTVs super period of the TIC\,279401253 pair of exoplanets is of the order of the libration frequency of the resonance angles $\theta_{1,2}$, i.e, $\sim$ 45 yr. Our estimate shows that the semiamplitude within this secular time scale in many cases is of the order of ten days. Thus, we could not give a meaningful prediction for the TTVs.

Nevertheless, the transiting planet is a prospective target for atmospheric investigation with the {\em James Webb Space Telescope} (JWST) and ground-based extremely-large telescopes, given its bright host star, $V \approx$ 11.9\,mag. the relatively short orbital period, and pronounced eccentricity. We plan to follow up this target with more RVs in an attempt to overcome the dynamical degeneracy and predict TTVs for future investigations.

\section{Summary and Conclusions}
\label{sec4}

We report the discovery of a warm pair of giant planets around the G-dwarf star TIC\,279401253. The system is revealed by $\tess$ light curve photometry and precise Doppler spectroscopy with FEROS and HARPS. Using the coadded HARPS spectra, we derived a stellar mass of M$_\star$ = 1.13$_{-0.03}^{+0.02}$ M$_\odot$ and a stellar radius of R$_\star$ = 1.06$_{-0.01}^{+0.01}$ R$_\odot$, among other physical and atmospheric stellar parameters. Using these stellar mass and radius estimates, we extensively analyzed the available data and constructed orbital posterior probability distributions. As a next step, we thoroughly analyzed the planetary system's dynamical architecture.\looseness=-5  

\begin{figure}[tp]
\begin{center}$
\begin{array}{ccc}

\includegraphics[width=9.0cm]{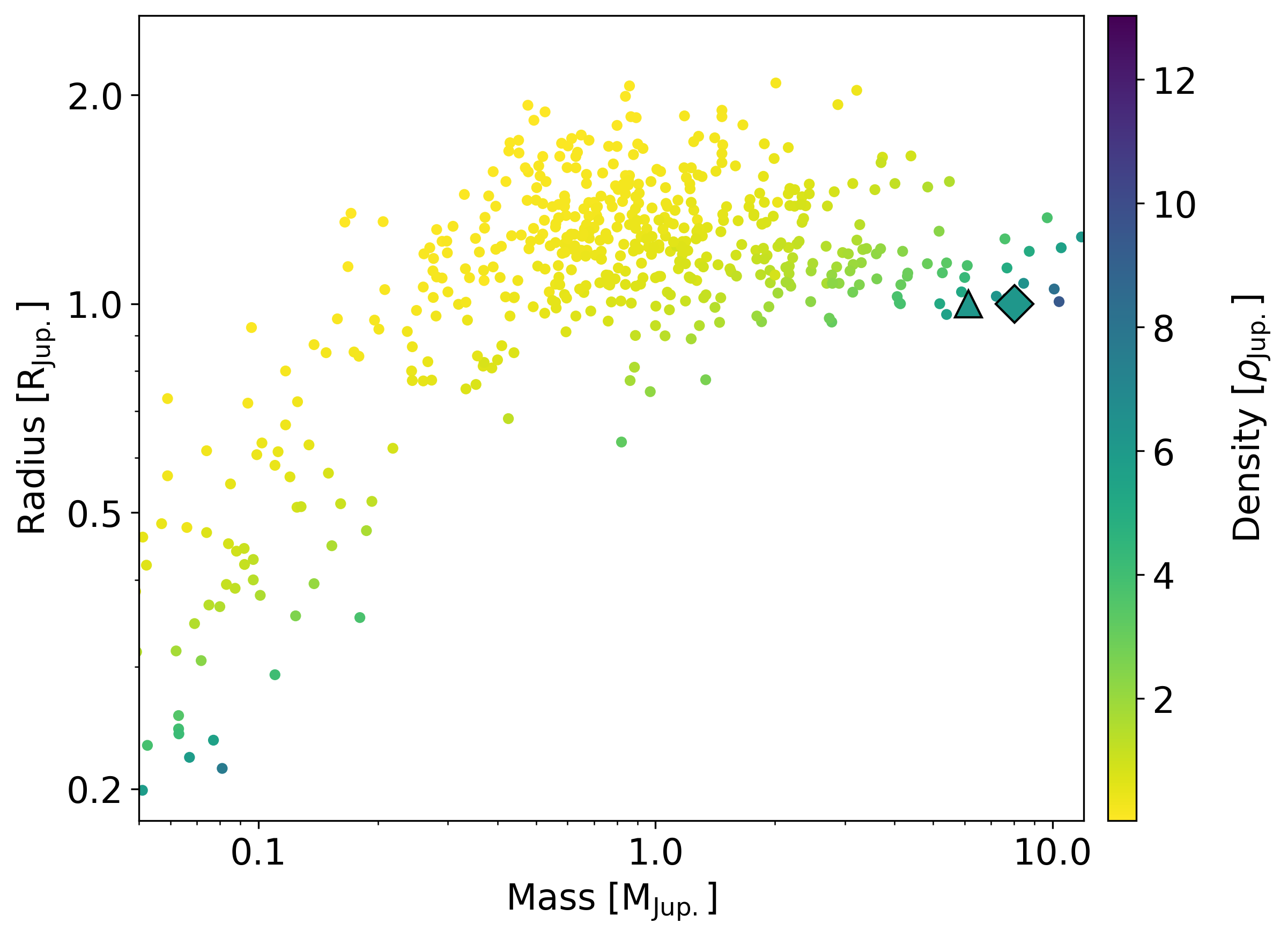}

 \end{array} $
\end{center}

\caption{Mass-radius distribution of exoplanets, colour-coded by their estimated mean density. TIC\,279401253\,b is marked with a triangle, among the densest warm Jovian planets found up-to-date. Assuming that TIC\,279401253\,c also has a Jupiter-like radius, its position is marked with diamond-shaped sign.\looseness=-5  
%its position will roughly match with that of TIC\,279401253\,b.
}
 
\label{RaMa} 
\end{figure}

 TIC\,279401253 b is a transiting massive-Jovian planet with a measured mass of m$_{\rm b}$ = 6.14$_{-0.42}^{+0.39}$$M_{\rm jup}$, and radius of $R_{\rm b}$ = 1.00$_{-0.04}^{+0.04}$ $R_{\rm jup}$. Thus, the estimated density of TIC\,279401253 b is $\rho_{\rm b}$ = 8.2$_{-1.1}^{+1.1}$ g\,cm$^{-3}$. \autoref{RaMa} shows the distribution of planets with measured radius and mass, colour-coded for their density. The position of TIC\,279401253\,b on this plot places it among the densest planets discovered so far.

 Based on the available RV data, we conclude that the outer planet, TIC\,279401253\,c, is similar to the inner super-Jovian planet with a minimum mass of $m_{\rm c}$ = 8.02$_{-0.18}^{+0.18}$ $M_{\rm jup}$. 
 The available $\tess$ light curve from Sectors 4 and 31 do not have sufficient coverage to reveal whether TIC\,279401253\,c is transiting, but assuming its radius is consistent with that of TIC\,279401253 b, then both planets in the system have high densities.  
 
 The warm pair of massive planets is found at the 2:1 period ratio commensurability with orbital periods of $P_{\rm b}$ =  76.80$_{-0.06}^{+0.06}$\,d for the inner planet, and $P_{\rm c}$ = 155.3$_{-0.7}^{+0.7}$\,d for the outer, respectively.
We performed detailed N-body simulations of the posterior probability distributions to reveal the long-term stability and the overall dynamics of the TIC\,279401253 system. We found that 73.4\% of the posterior samples are stable for 1\,Myr, and potentially beyond. The evolution of the apsidal alignment angle $\Delta\omega$, and the characteristic 2:1 MMR angles $\theta_1$ and $\theta_2$, exhibit libration about $0^\circ$. Thus, our numerical orbital analysis of the system unveiled that the massive pair of planets are deeply inside the first-order 2:1 MMR.\looseness=-4

The TIC\,279401253 system is strikingly similar to the  2:1 MMR pair orbiting HD\,82943, as reported by \citet{Tan2013}. Using a self-consistent dynamical fitting to the RV data of HD\,82943, \citet{Tan2013} unveiled two massive Jovian planets whose orbital geometry, physical, and dynamical characteristics are analogous to those of TIC\,279401253. For instance, \citet{Tan2013} found that the HD\,82943 system is very likely inclined to $i$ = 19.5$\pm$5 deg, which makes the dynamical masses of the HD\,82943 planets consistent with those of TIC\,279401253\,b, and c (see their Table 6). Furthermore, the orbital geometry and dynamical evolution of the two systems inside the 2:1 MMR exhibit practically the same pattern \citep[see our \autoref{evol_plot}, and, e.g., Figure\,7 in][]{Tan2013}.

It is plausible to assume that both systems have undergone a common formation and orbital evolution history. Such massive planet pairs are likely assembled during convergent migration in massive primordial circumstellar disks \citep[e.g.,][]{Lee2002, Kley2012}.
For instance, such massive planets must have gone through a slow type II migration. Given the almost equal mass ratio and large osculating eccentricities, it suggests that the planets were locked in the 2:1 MMR with initially nonzero eccentricity \citep[see,][]{Lee2004}. This is intriguing since planet-disk interactions tend to damp the eccentricity. Nonetheless, \citet{Papaloizou2001} showed that planet-disk interactions could pump the eccentricity of very massive Jovians up to $\approx$ 0.25, which is a plausible mechanism for eccentricity excitation in the disk. Alternatively, the migration of massive planets through the disk creates wide gaps, and this leads to excitation of the eccentricities in the disk \citep[see, e.g.,][]{GOT80, 2003ApJ...585.1024G}. 

Therefore, the TIC\,279401253 and HD\,82943 systems and their dynamical similarities are crucial forensic evidence for planetary formation mechanisms.

\begin{acknowledgements}

This research has made use of the Exoplanet Follow-up Observation Program website, which is operated by the California Institute of Technology, under contract with the National Aeronautics and Space Administration under the Exoplanet Exploration Program.
Funding for the $\tess$ mission is provided by NASA's Science Mission directorate.
This paper includes data collected by the $\tess$ mission, which are publicly available from the Mikulski Archive for Space Telescopes (MAST).
Resources supporting this work were provided by the NASA High-End Computing (HEC) Program through the NASA Advanced Supercomputing (NAS) Division at Ames Research Center for the production of the SPOC data products.
V.B., D.A., D.S., M.M., and T.T.\ acknowledge support by the BNSF program "VIHREN-2021" project No. KP-06-DV-5/15.12.2021. T.T. acknowledges support by the DFG Research Unit FOR 2544 "Blue Planets around Red Stars" project No. KU 3625/2-1. A.J., R.B., M.H.\ and F.R.\ acknowledge support from ANID -- Millennium  Science  Initiative -- ICN12\_009. D.D.\ acknowledges support from the NASA Exoplanet Research Program grant 18-2XRP18\_2-0136. A.J.\ acknowledges additional support from FONDECYT project 1210718. R.B.\ acknowledges support from FONDECYT project 11200751.  This work was also funded by the Data Observatory Foundation. We thank Yair Judkovsky for useful discussion on the TTV analysis. We thank the anonymous reviewer for very helpful comments and suggestions.
 
\end{acknowledgements}

 \newpage
 
\bibliographystyle{aasjournal}
\bibliography{bibliography}

\begin{appendix} %First online appendix

\label{appendix}

In this Appendix, \autoref{Kep_cornerplot} shows the posterior probability distribution of the joint Doppler and $\tess$ photometry modeling with {\tt Exo-Striker}, in \autoref{table:FEROS_RVs} and \autoref{table:HARPS_RVs} we list the spectroscopically derived RVs and activity index time-series from FEROS, and HARPS, respectively.

 \setcounter{table}{0}
\renewcommand{\thetable}{A\arabic{table}}

\setcounter{figure}{0}
\renewcommand{\thefigure}{A\arabic{figure}}

\begin{figure*}[tp]
\begin{center}$
\begin{array}{ccc} 

\includegraphics[width=18.0cm]{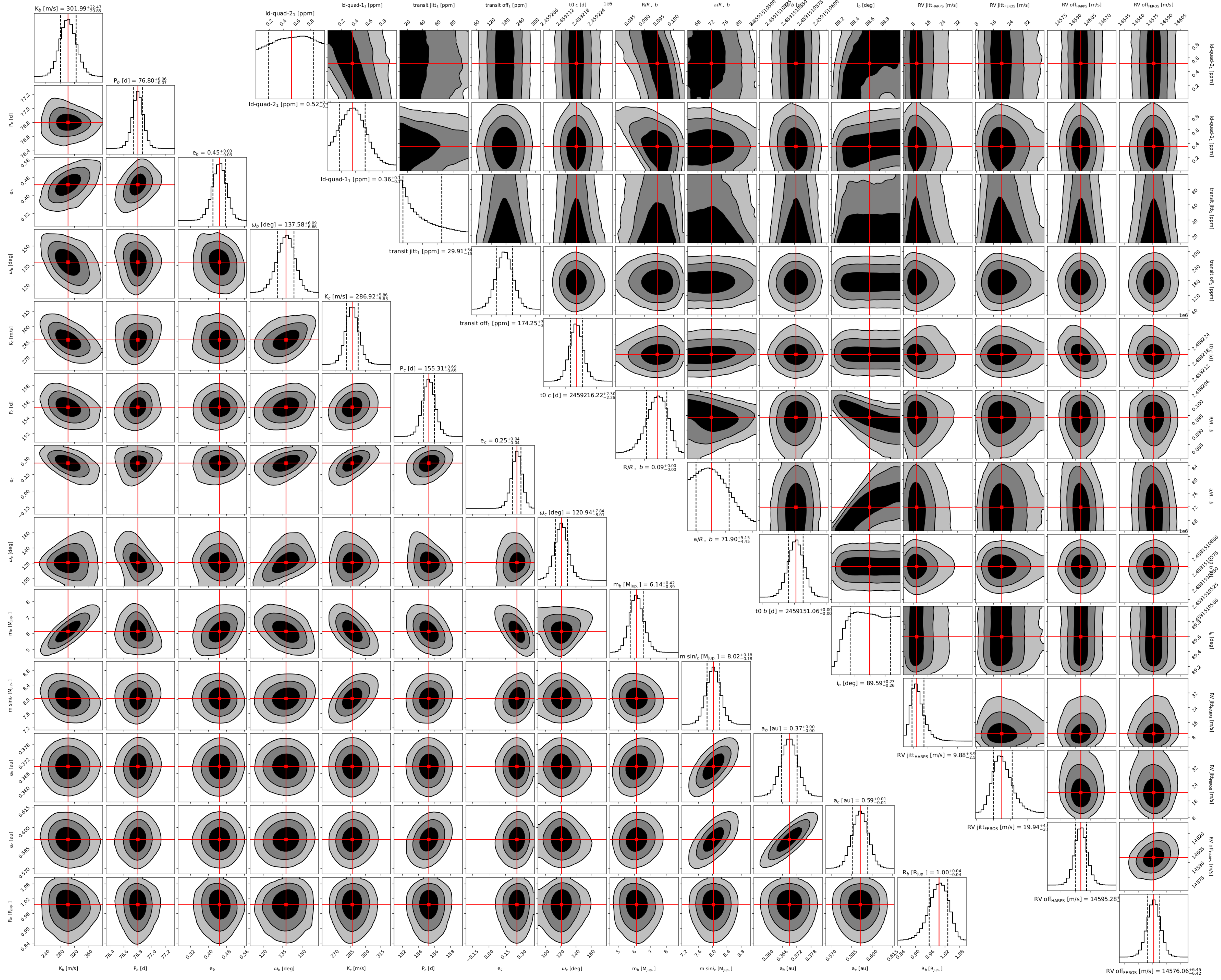} 

\end{array} $
\end{center}

\caption{Nested Sampling posteriors the two-planet system TIC\,279401253. 
}
 
\label{Kep_cornerplot} 
\end{figure*}

\begin{table}
\caption{FEROS Doppler and activity measurements of TIC279401253} 
\label{table:FEROS_RVs} 

\centering  

\begin{tabular}{c c c c c} 

\hline\hline    
\noalign{\vskip 0.5mm}

Epoch [JD] & RV [m\,s$^{-1}$] & $\sigma_{\rm RV}$ [m\,s$^{-1}$] & BIS [m\,s$^{-1}$] & $\sigma_{\rm BIS}$ [m\,s$^{-1}$]  \\  

\hline     
\noalign{\vskip 0.5mm}    

2459270.543   &   14574.3   &    7.6 & 7.0   &    10.0        \\ 
2459273.545   &   14643.0   &    9.6  & -23.0   &    12.0        \\ 
2459276.533   &   14660.3   &    8.8  & -29.0   &    12.0        \\ 
2459279.508   &   14710.4   &    8.2 &  -5.0   &    11.0        \\ 
2459281.511   &   14727.6   &    7.4 &  13.0   &    10.0        \\ 
2459284.548   &   14758.1   &    8.8  & -16.0   &    12.0        \\ 
2459483.716   &   14768.8   &    7.4 &  20.0   &    10.0        \\ 
2459485.732   &   14794.3   &    9.2 &  5.0   &    12.0        \\ 
2459493.858   &   14913.7   &    9.2  & 9.0   &    12.0        \\ 
2459505.706   &   14946.1   &    7.4  & -26.0   &    10.0        \\ 
2459506.738   &   14932.5   &    7.3  & -14.0   &    10.0        \\ 
2459515.848   &   14917.1   &    7.3 &  22.0   &    10.0        \\ 
2459542.720   &   13892.6   &    11.1  & 15.0   &    14.0        \\ 
2459548.660   &   14039.0   &    7.8 &  -6.0   &    11.0        \\ 
2459587.626   &   14689.6   &    7.2  & 0.0   &    10.0        \\ 
2459644.531   &   14878.9   &    9.8 &  20.0   &    13.0        \\ 
2459657.509   &   14937.9   &    8.2  & -10.0   &    11.0        \\ 
2459860.777   &   14075.8   &    8.0  & 13.0   &    11.0        \\ 
2459862.698   &   14091.6   &    9.4  & 35.0    &    12.0         \\ 
  
\hline           
\end{tabular}

\end{table}

\begin{table}
\caption{HARPS-DRS Doppler and activity measurements of TIC279401253  } 
\label{table:HARPS_RVs} 

\centering  

\begin{tabular}{c c c c c c} 

\hline\hline    
\noalign{\vskip 0.5mm}

Epoch [JD] & RV [m\,s$^{-1}$] & $\sigma_{RV}$ [m\,s$^{-1}$] & FWHM [m\,s$^{-1}$] &  BIS [m\,s$^{-1}$] & Contrast   \\  

\hline     
\noalign{\vskip 0.5mm}    

2459471.788   &   14509.7   &    2.8    &   7606.6   &    -1.9  &   45.9       \\ 
2459472.833   &   14544.4   &    2.9    &   7626.7   &    5.9  &   45.8       \\ 
2459492.795   &   14922.7   &    3.3    &   7608.6   &    0.4  &   45.8       \\ 
2459504.640   &   14971.6   &    3.5    &   7594.8   &    -23.0  &   45.8       \\ 
2459517.727   &   14921.4   &    2.6    &   7578.6   &    -0.4  &   45.9       \\ 
2459534.741   &   14328.6   &    2.2    &   7578.3   &    9.9  &   45.9       \\ 
2459561.786   &   14293.7   &    4.1    &   7591.9   &    11.7  &   46.1       \\ 
2459572.746   &   14489.2   &    2.6    &   7579.1   &    31.5  &   46.3       \\ 
2459625.614   &   14505.3   &    3.5    &   7523.2   &    -39.3  &   45.5       \\ 
2459640.574   &   14826.0   &    3.0    &   7584.4   &    -19.9  &   46.1       \\ 
2459642.559   &   14854.1   &    2.6    &   7595.5   &    4.9  &   46.0       \\ 
2459661.514   &   14970.8   &    4.4    &   7623.1   &    -34.9  &   46.2       \\ 
2459859.724   &   14091.4   &    3.2    &   7617.2   &    9.8  &   45.8       \\ 
2459867.729   &   14240.4   &    2.6    &   7592.3   &    -5.3  &   46.0       \\ 
  
\hline           
\end{tabular}

\end{table}

\end{appendix}

\end{document}